\documentstyle[twocolumn]{article}
\input psfig
\begin{document}
\title{Predictions of the Gamma-Ray Burst model of Ultra High
Energy Cosmic Rays}
\author{Eli Waxman\\
\it Institute for Advanced Study, Princeton, NJ 08540}
\maketitle
\begin{abstract}
The cosmological gamma-ray burst (GRB) model for the production of ultra-
high energy cosmic rays is described, and the current
observational evidence which support it discussed. Several predictions
of the model are presented, which would allow it to be tested by
future high energy cosmic ray 
and gamma-ray experiments. If the predicted signatures of the
GRB model are observed, they will not only corroborate the model,
but will also provide information about the source population, and will
allow to investigate the 
unknown structure of the inter-galactic magnetic field.
\end{abstract}

\section{Introduction}

Recent gamma ray and cosmic ray observations give increasing evidence
that the sources of gamma ray bursts (GRBs) and of cosmic rays with 
energy $E>10^{19}{\rm eV}$ are cosmological (see \cite{cos} 
for GRBs; [2--5] for cosmic rays).
The sources of both phenomena, however, remain unknown. In particular, 
most of the cosmic ray 
sources discussed so far have difficulties in accelerating
cosmic rays up to the highest observed energies \cite{Hillas}.
Furthermore, the arrival directions of the few cosmic rays detected above
$10^{20}{\rm eV}$ are inconsistent with the position of any astrophysical 
object that is likely to produce high energy particles \cite{obj}, since
the distance traveled by such particles must be smaller
than $100{\rm Mpc}$ \cite{dist}.

Although the source of GRBs is unknown, their observational
characteristics impose strong constraints on the physical conditions
in the $\gamma$-ray emitting region \cite{scenario},
which imply that protons may be accelerated by Fermi's mechanism 
in this 
region to energies $10^{20} - 10^{21}{\rm eV}$ \cite{Wa,Vietri}.
The observed energy spectrum of cosmic rays above $10^{19}
{\rm eV}$ (UHECRs) is consistent with a cosmological distribution of
sources of protons with a power-law generation spectrum typical of
Fermi acceleration \cite{Wb}. Furthermore,
the average rate (over volume and time) at which energy
is emitted as $\gamma$-rays by GRBs and in UHECRs
in the cosmological scenario is, remarkably, comparable 
\cite{Wa,Wb}. These facts suggest that GRBs and UHECRs 
have a common origin.

We describe the GRB model for UHECRs in Sec. \ref{subsec:fermi}, 
and discuss the current
observational evidence which support it 
in Sec. \ref{subsec:spec}. 
Several predictions of the
model are presented In Sec. \ref{sec:pred}. Sec. \ref{sec:conc} contains
a discussion.

\section{The GRB model}
\label{sec:model}

\subsection{Fermi acceleration in dissipative wind models of GRB's}
\label{subsec:fermi}

Whatever the ultimate source of GRBs is, observations strongly suggest the 
following similar scenario for the production of the bursts \cite{scenario}. 
The rapid rise time, $\sim{\rm ms}$, observed in some bursts implies
that the sources are compact, with a linear scale $r_0\sim10^7{\rm cm}$. The
high luminosity required for cosmological bursts, 
$L\sim10^{51}{\rm erg}{\rm s}^{-1}$,
then results in an initially optically thick (to pair creation) plasma
of photons, electrons and positrons, which expands and accelerates to 
relativistic velocities. In fact, the hardness of the observed
photon spectra, which extends to $\sim100{\rm MeV}$, implies 
that the $\gamma$-ray emitting region must be moving relativistically, 
with a Lorentz factor $\gamma$ of order a few hundreds.

If the observed radiation is due
to photons escaping the expanding ``wind'' as it becomes optically thin, two 
problems arise. First, the photon spectrum is quasi-thermal,
in contrast with observations. Second, the plasma is expected to be ``loaded''
with baryons which may be injected with the radiation or present in the 
atmosphere surrounding the source. A small baryonic load, $\geq10^{-8}
{M_\odot}$, increases the optical depth (due to Thomson scattering) so that 
most of the radiation energy is converted to
kinetic energy of the relativistically expanding baryons before the plasma
becomes optically thin. 
To overcome both problems, it was suggested that the
observed burst is produced once the kinetic energy of the ultra-relativistic 
ejecta is dissipated, due to collision of the relativistic baryons
with the inter-stellar medium or due to internal collisions within the ejecta,
at large radius $r=r_d>10^{12}{\rm cm}$ beyond
the Thomson photosphere, and then radiated as $\gamma$-rays \cite{coll}.

Since $\gamma\gg1$, substantial dissipation of
kinetic energy at $r=r_d$
implies that the random motions in the wind rest frame are (at least mildly)
relativistic. The relativistic random motions 
are likely to give rise to a turbulent build up of magnetic
fields, and therefore to Fermi acceleration
of charged particles. We derive below the constraints that should be satisfied
by the wind parameters in order to allow acceleration of protons at the
dissipation region to $\sim10^{20}{\rm eV}$.

The most restrictive requirement, which rules out the possibility of 
accelerating 
particles to energies $\sim10^{20}{\rm eV}$ in most astrophysical objects, 
is that
the particle Larmor radius $R_L$ should be smaller than the system size
\cite{Hillas}. In our scenario we must apply a more stringent requirement.
Due to the wind expansion the internal energy is decreasing and therefore
available for proton
acceleration (as well as for $\gamma$-ray production) only
over a comoving time $t_d\sim r_d/\gamma c$. The typical Fermi
acceleration time 
is $t_a\simeq R_L/c$ \cite{Hillas}, 
leading to the requirement $R_L<r_d/\gamma$.
This condition sets a lower limit for the required
comoving magnetic field strength \cite{Wa}, 
\begin{equation}
\left({B\over B_{e.p.}}\right)^2>0.15\gamma_{300}^2
E_{20}^2L_{51}^{-1},\label{larmor}
\end{equation}
where $E=10^{20}E_{20}{\rm eV}$, $\gamma=300\gamma_{300}$, 
$L=10^{51}L_{51}{\rm erg}{\rm\ s}^{-1}$
is the wind luminosity, and $B_{e.p.}$ is the equipartition field,
i.e. a field with
comoving energy density similar to that associated with the random
energy of the baryons.

The accelerated proton energy is also limited by energy loss
due to synchrotron radiation. 
The condition that the synchrotron loss time
should be smaller than the acceleration time is
\begin{equation}
B<3\times10^5\gamma_{300}^{2}E_{20}^{-2}{\rm G}.\label{sync}
\end{equation}
Since the equipartition field is inversely proportional to the radius $r$,
this condition may be satisfied simultaneously with (\ref{larmor}) provided
that the dissipation radius is large enough, i.e.
\begin{equation}
r_d>10^{12}\gamma_{300}^{-2}E_{20}^3{\rm cm}.\label{dis}
\end{equation}
The high energy protons lose energy also in interaction 
with the wind photons (mainly through pion production). It can be 
shown, however, that this energy loss is less important than the synchrotron 
energy loss \cite{Wa}.

The conditions (\ref{larmor}-\ref{dis}) imply, that
a dissipative ultra-relativistic wind,
with luminosity and bulk Lorentz factor implied by GRB observations,
satisfies the constraints necessary to allow the acceleration of protons 
to energies $\sim10^{20}{\rm eV}$ by second order Fermi acceleration,
provided that turbulent build up
of magnetic fields during dissipation gives rise to fields which are
close to equipartition. It should be noted that equipartition field is
also required in most of the dissipative relativistic wind models for GRBs.
Gamma-rays are produced in these models by synchrotron and synchro-self Compton
radiation of relativistic electrons produced by the dissipation shocks. 
Equipartition field is required in this case to allow efficient radiation
of the electrons \cite{scenario}.

Although the details of the mechanism
of turbulent build up of an equipartition field are not fully understood, it
seems to operate in a variety of astrophysical systems. 
We note that a magnetic field of
this strength may exist in the plasma prior to the onset of internal collisions
if a substantial part of the wind luminosity is initially, i.e. at $r\sim r_0$,
provided by magnetic field energy density,
and if the field becomes
predominantly transverse. The pre-existing field may suppress small scale
turbulent motions. 
In this case shocks coherent over a scale $R\sim r_d/\gamma$
would exist and protons would be accelerated by first rather than second
order Fermi mechanism. The constraints (\ref{larmor}-\ref{dis}) are
valid in this case too, therefore leaving the above conclusions unchanged.

\subsection{UHECR spectrum and flux}
\label{subsec:spec}

\begin{figure}[t]
\centerline{\psfig{figure=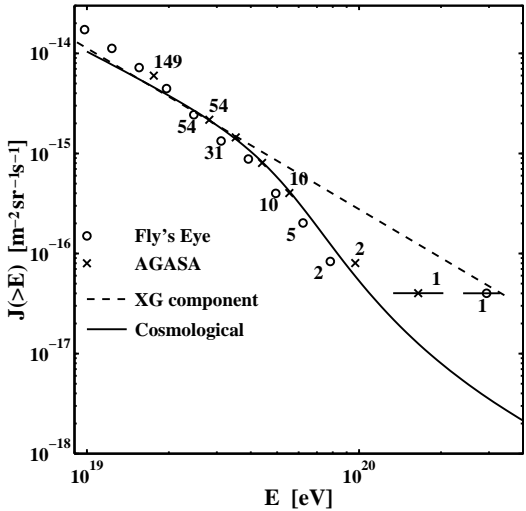,width=2.7in}}
\caption{
The UHECR flux expected in a cosmological model, compared to the Fly's Eye 
and AGASA data. Integers indicate the number of events observed.
$1\sigma$ energy error bars are shown for the highest energy events. 
The dashed line denotes the fit by
Bird {\it et al.} \cite{Fly} for the extra-galactic flux.}
\label{fig1}
\end{figure}

In the GRB model for UHECR production described above, 
the high energy cosmic rays are 
protons accelerated by Fermi's mechanism 
in sources that are distributed throughout
the universe. 
In Fig.\ \ref{fig1} we compare the UHECR spectrum,
reported by the Fly's Eye and the AGASA experiments \cite{Fly,AGASA2}, 
with that expected from 
a homogeneous cosmological distribution of sources, each generating
a power law differential spectrum of high energy protons
$dN/dE\propto E^{-2.2}$, as typically
expected from Fermi acceleration (e.g. \cite{Hillas}). 
(For this calculation we have used a flat universe
with zero cosmological constant and $H_0=75{\rm km}\ {\rm s}^{-1}$; The 
spectrum is insensitive to the cosmological parameters and to source
evolution, since most of the cosmic rays arrive from distances 
$<500{\rm Mpc}$). The AGASA flux at $3-10\times10^{18}{\rm eV}$
is $\sim1.7$ times higher than that reported by the Fly's Eye, corresponding
to a systematic $\sim20\%$ larger estimate of event energies in the AGASA
experiment compared to the Fly's Eye experiment (see also \cite{Fly,AGASA}).
We have therefore multiplied in Fig.\ \ref{fig1} the Fly's Eye 
energy by $1.1$ and the AGASA energy by $0.9$. Bird {\it et al.} 
\cite{Fly} find that the Fly's Eye flux in the energy range $4\times10^{17}
-4\times10^{19}{\rm eV}$ can be fitted by a sum of two power laws: A
steeper Galactic component with $J\propto E^{-2.5}$ dominating at lower
energy, and a shallower extra-Galactic component with $J\propto E^{-1.6}$ 
dominating at higher energy. 
The Bird {\it et al.} fit to the extra-Galactic component is also 
shown in Fig.\ \ref{fig1}. 

The data are consistent with the cosmological model for $E>2\times10^{19}
{\rm eV}$. Furthermore, the flux predicted by the model
at lower energy is consistent with the Bird 
{\it et al.} fit to the extra-Galactic component.
(The flux deduced from the highest energy
event in the Fly's Eye data is significantly
higher than that predicted from the cosmological model. However, 
the statistical significance of the apparent discrepancy
is not high: For the Fly's Eye exposure, the model predicts an average of
$\sim1.3$ events above $10^{20}{\rm eV}$, and the probability that the
first event observed at this energy range is above $2\times10^{20}{\rm eV}$
is $\sim15\%$.) 
The deficit in the number of events detected above 
$5\times10^{19}{\rm eV}$, compared to a
power-law extrapolation of the flux at lower energy,
is consistent with that expected due
to a cosmological ``black-body cutoff''. However, with current data the
``cutoff'' is detected with only $2\sigma$ significance. 

The present rate at which energy should be produced as $10^{19}$--$10^{20}
{\rm eV}$ protons by the cosmological cosmic ray sources in order to 
produce the observed flux is $4\pm2\times10^{44}{\rm erg\ Mpc}^{-3}
{\rm yr}^{-1}$. This rate is comparable to that produced in $\gamma$-rays
by cosmological GRBs: The rate of cosmological GRB events is $\nu_\gamma\simeq
3\times10^{-8}{\rm Mpc}^{-3}{\rm yr}^{-1}$ \cite{rate}, each producing
$\approx3\times10^{51}{\rm erg}$, corresponding to a $\gamma$-ray energy
production rate of $\sim10^{44}{\rm erg\ Mpc}^{-3}
{\rm yr}^{-1}$.

The above analysis implies that the GRB model of UHECR production would
produce UHECR flux consistent with that observed, provided the efficiency
with which the wind kinetic energy is converted to $\gamma$-ray and UHECR 
energy is similar. There is, however, one additional point which requires
consideration. The energy of the most
energetic cosmic ray detected by the Fly's Eye experiment is in excess of
$2\times10^{20}{\rm eV}$, and that of the most
energetic AGASA event is above $10^{20}{\rm eV}$. On a
cosmological scale, the distance traveled by such energetic particles is
small: $<100{\rm Mpc}$ ($50{\rm Mpc}$) for the AGASA (Fly's Eye) event
(e.g., \cite{dist}). Thus, the detection of these events over a $\sim5
{\rm yr}$ period can be reconciled with the rate of nearby GRBs, $\sim1$
per $50\, {\rm yr}$ in the field of view of the CR experiments out to $100
{\rm Mpc}$ in a standard cosmological scenario \cite{rate}, only if
there is a large dispersion, $\geq50{\rm yr}$, in the arrival time of protons 
produced in a single burst (This implies that if a direct 
correlation between 
high energy CR events and GRBs, as recently suggested in
\cite{MU}, is observed
on a $\sim10{\rm yr}$ time scale, it would be strong evidence {\it against} a 
cosmological GRB hypothesis). 

The required dispersion
is likely to occur due to the combined effects of deflection 
by random magnetic fields and energy dispersion of the particles. 
Consider a proton of energy $E$ propagating through a magnetic field of 
strength $B$ and correlation length
$\lambda$. As it travels a distance $\lambda$, the proton is typically 
deflected by an angle $\alpha\sim\lambda/
R_L$, where $R_L=E/eB$ is the Larmor radius. The
typical deflection angle for propagation over a distance $d$ is
$\theta_s\sim(d/\lambda)^{1/2}\lambda/R_L$. This deflection results in a time
delay, compared to propagation along a straight line, of order
$\tau(E,d)\approx\theta_s^2d/c\approx(eBd/E)^2\lambda/c$.
The random energy loss UHECRs suffer as they propagate, owing to the 
production of pions, implies that 
at any distance from the observer there is some finite spread
in the energies of UHECRs that are observed with a given fixed energy.
For protons with energies
$>10^{20}{\rm eV}$ the fractional RMS energy spread is of order unity
over propagation distances in the range $10-100{\rm Mpc}$ (e.g. \cite{dist}).
Since the time delay is sensitive to the particle energy, this implies that
the spread in arrival time of UHECRs with given observed energy is comparable
to the average time delay at that energy $\tau(E,d)$. The 
field required to produce $\tau>100{\rm yr}$ is
\begin{equation}
B\sqrt{\lambda}>10^{-11}E_{20}d_{100}^{-1}{\rm G\ Mpc}^{1/2},
\label{Bmin}
\end{equation}
where $d=100d_{100}{\rm Mpc}$. 
The required field is consistent with
the current upper limit for the inter-galactic magnetic 
field, $B\lambda^{1/2}\le10^{-9}{\rm G\ Mpc}^{1/2}$ \cite{field}.
A time broadening over $\tau\gg100{\rm yr}$
is therefore possible.

It should be noted, that a GRB producing $3\times10^{51}{\rm erg}$ in
UHECRs at $50{\rm Mpc}$ distance, would produce a total fluence at Earth
of $\sim2$ cosmic rays above $10^{19}{\rm eV}$ per ${\rm km}^2$.
In the presence of a magnetic field induced time delay,
the typical distance $d_m(E)$ to the brightest source observed
over an energy range $\Delta E$ around $E$, with $\Delta E/E\sim 1$,
is the radius of a sphere within which the average time between bursts is
equal to the characteristic time delay $\tau[E,d_m(E)]$;
i.e. $4\pi d_m^3\nu_\gamma\tau(E,d_m)/3=1$. 
Thus, the brightest source distance is
\begin{equation}
d_m(E)\simeq 30\left({B\sqrt{\lambda}\over10^{-11}{\rm G\ Mpc}^{1/2}}
\right)^{-2/5}E_{19}^{2/5}\,{\rm Mpc},
\label{Dm}
\end{equation}
and its flux is $f\sim0.1E_{19}^{-3/5}(B\lambda^{1/2}/10^{-11}
{\rm G\ Mpc}^{1/2})^{-2/5}$ per ${\rm km}^{2}{\rm yr}$ \cite{Jordi2}. 
Here $E=10^{19}E_{\rm 19}{\rm eV}$.

\section{Predictions}
\label{sec:pred}

\subsection{The Number and Spectra of Bright Sources}
\label{subsec:bright}

The initial proton energy, necessary to have an observed energy $E$,
increases with source distance due to propagation energy losses.
The rapid increase of the initial energy after it exceeds, due to
electron-positron production, the threshold for pion production effectively
results in a cutoff distance, $d_c(E)$, beyond which sources do not contribute
to the flux above $E$. Since $d_c(E)$ is a decreasing function of $E$, for
a given number density of sources there is a critical energy $E_c$, above which
only one source (on average) contributes to the flux. 
For bursting sources, $E_c$ depends on the product of the burst rate $\nu$
and the time delay. In the GRB model, the burst rate is given by the GRB
rate $\nu=\nu_\gamma$, which is determined from the GRB flux distribution.
The time delay depends on the unknown properties of 
the intergalactic magnetic field, $\tau\propto B^2\lambda(d/E)^2$. 
However, the rapid
decrease of $d_c(E)$ with energy near $10^{20}{\rm eV}$ implies that
$E_c$ is not very sensitive to the unknown value of $B^2\lambda$. 
For the range allowed
for the GRB model, $10^{-11}{\rm G\ Mpc}^{1/2}\le B\lambda^{1/2}\le10^{-9}
{\rm G\ Mpc}^{1/2}$ (the lower
limit determined by (\ref{Bmin}), and the upper limit by Faraday rotation 
observations \cite{field}),
the allowed range of $E_c$ is 
$10^{20}{\rm eV}\le E_c\le3\times10^{20}{\rm eV}$.

\begin{figure}[t]
\centerline{\psfig{figure=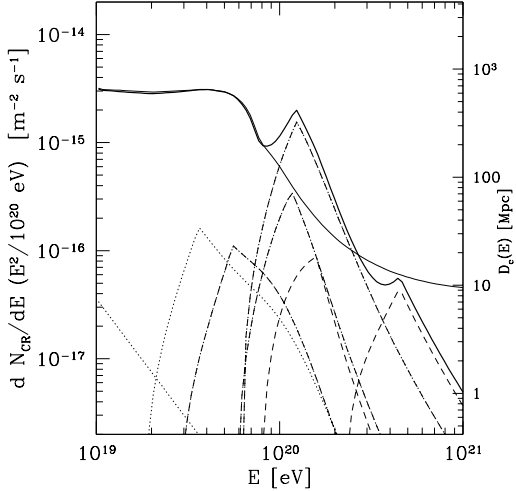,width=2.7in}}
\caption{Results of a Monte-Carlo realization of the bursting sources
model: Thick solid line- overall 
spectrum in the realization;
Thin solid line- average spectrum, this
curve also gives $d_c(E)$;
Dotted lines- spectra of brightest sources at different energies.
}
\label{figNc}
\end{figure}

Fig. \ref{figNc} presents the flux obtained in one realization of
a Monte-Carlo simulation described in ref. \cite{Jordi1} of the total
number of UHECRs received from GRBs at some fixed time. 
For each
realization the positions (distances from Earth) and
times at which cosmological GRBs occurred were randomly drawn, 
assuming an average rate $\nu_\gamma=
2.3\times10^{-8}{\rm Mpc}^{-3}{\rm yr}^{-1}$, an intrinsic
generation spectrum $n_p(E) \propto E^{-2}{\rm d}E$, and $E_c
=1.4\times10^{20}{\rm eV}$. 
Most of the realizations gave an overall spectrum similar to that obtained
in the realization of Fig. \ref{figNc} when the brightest source of this 
realization (dominating at $10^{20}{\rm eV}$) is not included.
At $E < E_c$,
the number of sources contributing to the flux is very large, 
and the overall UHECR flux received at any
given time is near the average (the average flux is that obtained when 
the UHECR emissivity is spatially uniform and time independent).
At $E > E_c$, the flux will generally be much lower than the average,
because there will be no burst within a distance $d_c(E)$ having taken
place sufficiently recently. There is, however, a significant probability
to observe one source with a flux higher than the average.
A source similar to the brightest one in Fig. \ref{figNc}
appears $\sim5\%$ of the time. 

At any fixed time a given burst is observed in UHECRs only over a narrow
range of energy, because if
a burst is currently observed at some energy $E$ then UHECRs of much lower 
energy from this burst have not yet arrived,  while higher energy UHECRs
reached us mostly in the past. As mentioned above, for energies above the 
pion production threshold, 
$E\sim5\times10^{19}{\rm eV}$, the dispersion in arrival times of UHECRs
with fixed observed energy is comparable to the average delay at that
energy. This implies that
the spectral width $\Delta E$ of the source at a given time is of order
the average observed energy, $\Delta E\sim E$.
Thus, bursting UHECR sources should have narrowly peaked energy
spectra,
and the brightest sources should be different at different energies.
For steady state sources, on the other hand, the brightest
source at high energies should also be the brightest one at low
energies, its fractional contribution to the overall flux decreasing to
low energy only as $d_c(E)^{-1}$.

\subsection{Spectra of Sources at $E<4\times10^{19}{\rm eV}$}
\label{subsec:Blambda}

The detection of UHECRs 
above $10^{20}{\rm eV}$ imply that the brightest sources 
must lie at distances smaller than $100{\rm Mpc}$.
UHECRs with $E\le4\times10^{19}{\rm eV}$
from such bright sources will suffer energy loss only by pair production,
because at $E < 5\times 10^{19}$ eV
the mean-free-path for pion production interaction
(in which the fractional energy loss is $\sim10\%$) is larger than 
$1{\rm Gpc}$. Furthermore, the energy loss due to pair production 
over $100{\rm Mpc}$ propagation is only $\sim5\%$.

In the case where the typical displacement of the UHECRs 
due to deflections by inter-galactic magnetic fields is 
much smaller than the correlation length, $\lambda \gg d\theta_s(d,E)
\simeq d(d/\lambda)^{1/2}\lambda/R_L$,
all the UHECRs that arrive at the
observer are essentially deflected by the same magnetic field structures, 
and the absence of random energy loss during propagation implies that
all rays with a fixed observed energy would reach the observer with exactly
the same direction and time delay. At a fixed time, therefore, the source would
appear mono-energetic and point-like. In reality,
energy loss due to pair production
results in a finite but small spectral and angular width, 
$\Delta E/E\sim\delta\theta/\theta_s\le1\%$ \cite{Jordi2}.

In the case where the typical displacement of the UHECRs is 
much larger than the correlation length, $\lambda \ll d\theta_s(d,E)$,
the deflection of different UHECRs arriving at the observer
are essentially independent. Even in the absence of any energy loss there 
are many paths from the source to the observer for UHECRs of fixed energy $E$
that are emitted from the source at an angle 
$\theta\le\theta_s$ relative to the source-observer line of sight. Along
each of the paths, UHECRs are deflected by independent magnetic field 
structures. Thus, the source angular size would be of order $\theta_s$
and the spread in arrival times would be comparable to the characteristic 
delay $\tau$, leading to $\Delta E/E\sim1$ even when there are no random
energy losses. The observed spectral shape of a nearby ($d<100{\rm Mpc}$) 
bursting source of UHECRs at 
$E<4\times10^{19}{\rm eV}$
was derived for the case $\lambda \ll d\theta_s(d,E)$ in 
\cite{Jordi2}, and is given by
\begin{equation}
{dN\over dE}\propto \sum\limits_{n=1}^{\infty} (-1)^{n+1}\, n^2\,
\exp\left[ -{2n^2\pi^2 E^2\over E_0^2(t,d)} \right]\quad,
\label{flux}
\end{equation}
where $E_0(t,d)=de(2{B^2\lambda}/3ct)^{1/2}$.  
For this spectrum, the ratio of the 
RMS UHECR energy spread to the average energy is $30\%$

\begin{figure}[t]
\centerline{\psfig{figure=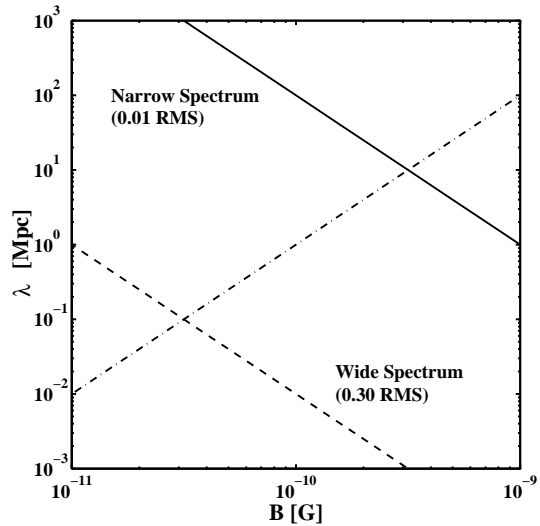,width=2.7in}}
\caption{The line $\theta_s d=\lambda$ for a source at 
$30{\rm Mpc}$ distance 
observed at energy $E\simeq10^{19}{\rm eV}$ (dot-dash line), shown with
the Faraday rotation upper limit $B\lambda^{1/2}
\le10^{-9}{\rm G\ Mpc}^{1/2}$ (solid line), and with the lower limit 
$B\lambda^{1/2}\ge10^{-11}{\rm G\ Mpc}^{1/2}$ required in the GRB model.
}
\label{figBL}
\end{figure}

Fig. \ref{figBL} shows the line $\theta_s d=\lambda$ in the $B-\lambda$ plane,
for a source at a distance $d=30{\rm Mpc}$
observed at energy $E\simeq10^{19}{\rm eV}$.
Since the $\theta_s d=\lambda$ line divides the
allowed region in the plane at $\lambda\sim1{\rm Mpc}$,
measuring the spectral width of bright sources would allow to determine
if the field correlation length is much larger, much smaller, or comparable
to $1{\rm Mpc}$.

\subsection{Correlation with Large Scale Structure (LSS)}
\label{subsec:LSS}

If the UHECR sources are indeed extra-Galactic, and
if they are correlated with luminous matter, then the 
inhomogeneity of the large scale galaxy distribution on 
scales $\le100{\rm Mpc}$ should be imprinted on the UHECR arrival directions.
The expected anisotropy signature
and its dependence on the relative bias 
between the UHECR sources and the galaxy 
distribution and on the (unknown) intrinsic UHECR source density
have been examined in \cite{Fisher}. The galaxy  
distribution was derived from the {\it IRAS} 1.2 Jy redshift
survey, which gives an acceptable description of the
LSS out to $\sim100{\rm Mpc}$ (see \cite{Fisher} for a detailed
analysis).

\begin{figure*}[t]
\vglue-0.8in
\centerline{\psfig{figure=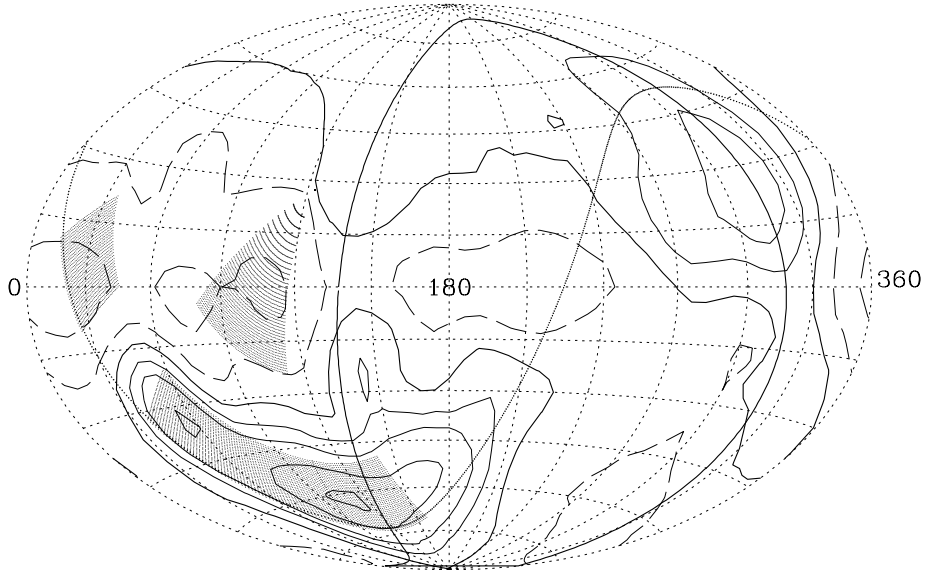,width=5in}}
\vglue-0.85in
\caption{Aitoff projection (Galactic coordinates) of the fractional 
deviation (from the all sky average) of the mean UHECR intensity.
The heavy curve denotes the zero contour. Solid (dashed) contours
denote positive (negative) fractional fluctuations at intervals
$[-0.5,\ -0.25,\ 0,\ 0.25,\ 0.50,\ 1.0,\ 1.5]$. The super-galactic
plane is denoted by the heavy solid curve roughly perpendicular
to the Galactic plane. The dotted curve denotes the 
Fly's Eye coverage of declination $>-10^\circ$. The shaded regions show
the high and low density regions used in the $X(E)$ statistic.
}
\label{figLSS}
\end{figure*}

Figures \ref{figLSS} 
presents a map of the angular dependence of the mean (over different 
realizations of source distribution) UHECR intensity, 
for $E\ge6\times10^{19}{\rm eV}$ and a model where UHECR sources
trace {\it IRAS} galaxies with no bias.
The map clearly reflects the inhomogeneity of the large-scale 
galaxy distribution- the overdense UHECR regions lie in the directions of 
the ``Great Attractor'' [composed of the Hydra-Centaurus 
($l=300$--$360^\circ$, $b=0$--$+45^\circ$)
and Pavo-Indus 
($l=320$--$360^\circ$, $b=-45$--$0^\circ$) super-clusters]
and the Perseus-Pisces super-cluster 
($l=120$--$160^\circ$, $b=-30$--$+30^\circ$).

In order to determine the
exposure required to discriminate between isotropic and LSS correlated
UHECR source distribution,
the distribution of a statistic similar in spirit to $\chi^2$ was considered, 
$X(E)=\sum_l {[n_l(E)- n_{l,I}(E)]^2/ n_{l,I}(E)}$,
where $n_l$ is the number of events detected in angular bin $l$ and 
$n_{l,I}$ is the average number expected for isotropic distribution (For 
the calculation of $X$ $24^\circ\times24^\circ$ bins were used; see also
Fig. \ref{figLSS}). From the $X(E)$ distributions, generated
by Monte-Carlo simulations of the UHECR source distributions, it was found
that the exposure required for a northern hemisphere detector to discriminate
between isotropic UHECR source distribution and an unbiased distribution that 
traces the LSS is approximately $10$ times the current Fly's Eye exposure
($0.1$ the expected Auger exposure).
If the UHECR source distribution is strongly biased, in a way similar to that
of radio galaxies, the required exposure
is $\sim3$ times smaller. Furthermore, with $10$ times the current 
Fly's Eye exposure, it would be possible to discriminate between biased
and non-biased source distribution.
The anisotropy signal is not sensitive to the 
currently unknown number density of sources. 

Stanev {\it et al.} \cite{Stanev} have recently reported that
the arrival directions of $E>4\times10^{19}{\rm eV}$ UHECR events detected
by the Haverah Park experiment show a concentration in the
direction of the super-galactic plane. In agreement with 
Stanev {\it et al.}, it is found in \cite{Fisher}
that the probability to obtain
the Haverah Park results assuming an isotropic source distribution is very
low. However, the results of \cite{Fisher} show that this probability is not
significantly higher for models where the source distribution traces 
the LSS; thus, the concentration of the Haverah Park events towards the 
super-galactic plane can not be explained by the known LSS. 
It is important to note that for the biased
model the probability to obtain the Haverah Park 
results is smaller than for the unbiased one. This reflects the fact that
the super-clusters, while concentrated towards the super-galactic plane, 
have offsets above and below the super-galactic plane which cause the inferred 
UHECR distribution to be less flattened than seen in the Haverah Park data.

\section{Conclusions}
\label{sec:conc}

The GRB model for UHECR production has several predictions, which
would allow it to be tested with future experiments.
In this model, the average number of sources 
contributing to the flux decreases with energy much more rapidly than in the 
case where the UHECR sources are steady \cite{Jordi1}. 
A critical energy 
exists, $10^{20}{\rm eV}\le E_c<3\times10^{20}{\rm eV}$,
above which a few sources produce most of the flux, and the 
observed spectra of these sources is narrow, $\Delta E/E\sim1$: 
the bright sources
at high energy should be absent in UHECRs of much
lower energy, since particles take longer to arrive the lower their
energy. In contrast, a model of steady sources predicts that the
brightest sources at high energies should also be the brightest ones at
low energies.

At the highest energies, where most of the cosmic rays should come only from 
a few sources, bursting sources should
be identified from only a small number of events from their coincident
directions. Many more cosmic rays need to be detected at lower energies, 
where many sources contribute to the flux, in order to identify sources.
Recently, the AGASA experiment reported the presence
of 3 pairs of UHECRs with angular separations (within each pair) 
$\le2.5^\circ$, roughly consistent with the measurement error,
among a total of 36 UHECRs with $E\ge4\times10^{19}{\rm eV}$ \cite{AGASA2}. 
The two highest energy AGASA events were in these pairs.
Given the total solid angle observed by the experiment, $\sim2\pi{\rm sr}$,
the probability to have found 3 pairs by chance is $\sim3\%$; and, given that
three pairs were found, the probability that the two highest energy events are
among the three pairs by chance is 2.4\%. Therefore, this observation favors
the bursting source model, although more data are needed to confirm it.

Above $E_c$, there is a significant probability to observe one 
source with a flux considerably higher than average. If such a source is 
present, its narrow spectrum may produce a ``gap'' in the overall 
spectrum, as demonstrated in Fig. \ref{figNc}. 
It had recently been argued \cite{TD} 
that the observation of such an energy gap would 
imply that the sources of $>10^{20}{\rm eV}$ UHECRs are different from
the sources at lower energy, hinting that these are 
produced by the decay of a new type of massive
particle. We see here that this is not the case when bursting sources
are allowed, owing to the time variability. If such an energy gap is
present, our model predicts that most of the UHECRs above the gap
should normally come from one source.
If our model is correct, then the Fly's Eye event above 
$2\times10^{20}{\rm eV}$ suggests that we live at one of the times 
when a bright source is present at high energies. However, 
given the present scarcity of UHECRs, no solid
conclusions can be drawn. With the projected Auger experiment 
\cite{huge}, the number of detected UHECRs would increase by
a factor $\sim 100$. If $E_c$ is $2\times10^{20}{\rm eV}$,
then a few bright sources above $10^{20}{\rm eV}$ should be identified.

For the GRB model, the expected number of events 
to be detected by the $5000{\rm km}^2$ Auger detectors from
individual bright sources at $E\sim10^{19}{\rm eV}$ is of order $100$
(cf. eq. (\ref{Dm}); see also \cite{Jordi2}). 
The spectral width of these sources depends on the correlation length
$\lambda$ of the inter-galactic magnetic field: Very narrow spectrum, 
$\Delta E/E\le1\%$, is expected for $\lambda>1{\rm Mpc}$, and a wider
spectrum, $\Delta E/E\sim1$, is expected for $\lambda\ll1{\rm Mpc}$
(see Fig. \ref{figBL}). With energy resolution of
$\sim10\%$, the Auger detectors would easily allow
to determine the spectral width of the sources, and therefore to put
interesting constraints on the unknown structure of the magnetic field.

If the distribution of UHECR sources traces the large scale 
structure (LSS) of luminous matter, 
large exposure detectors should clearly reveal
anisotropy in the arrival direction distribution of UHECRs above
$4\times10^{19}{\rm eV}$. With $10$ times the current Fly's Eye exposure
($0.1$ the expected Auger exposure), it would be possible to determine
whether the sources are distributed isotropically, or correlate with 
known LSS. Furthermore, it would be possible to determine whether or not 
the source
distribution is highly biased compared to {\it IRAS} galaxies (as radio 
galaxies are). Thus, the anisotropy signal would provide constraints on the
source population.

Finally, we would like to note another possible signature of the GRB model, 
which was not discussed above. 
The energy lost by the UHECRs as they propagate and interact with the 
microwave background is transformed by cascading into secondary GeV-TeV 
photons. A significant fraction of these photons can arrive with 
delays much smaller than the UHECR delay if much of inter-galactic space is
occupied by large-scale magnetic ``voids'', regions of size
$\ge5{\rm Mpc}$ and field weaker than $10^{-15}{\rm G}$. 
Such voids might be expected, for example, in models where a weak primordial 
field is amplified in shocked, turbulent regions of the intergalactic medium
during the formation of large-scale structure.
For a field strength $\sim4\times10^{-11}{\rm G}$ in the high field regions, 
the value required to account for observed galactic fields if the field were 
frozen in the protogalactic plasma, the delay of protons produced by a burst 
at a 
distance of $100{\rm Mpc}$ is $\sim100{\rm yr}$, and the fluence of secondary 
photons above $10{\rm GeV}$ on hour--day time scales is 
$I(>E)\sim10^{-6}E_{\rm TeV}^{-1}{\rm cm}^{-2}$ \cite{Paolo}. This fluence is 
close to the detection threshold of current high-energy $\gamma$-ray 
experiments.

\subsection*{Acknowledgments}
I would like to thank my collaborators in the work on which this article
is based, J. Miralda-Escud\'e, P. Coppi, T. Piran and K. Fisher, and Prof.
M. Nagano for the invitation to participate in the Tokyo International
Symposium on Extermely High Energy Cosmic Rays. This research
was partially supported by a W. M. Keck Foundation grant 
and NSF grant PHY 95-13835.

\end{document}